\documentstyle[prl,aps,preprint,tighten]{revtex}
\begin{document}
\newcommand{\as}{\tilde{\alpha}_0}
\newcommand{\ds}{\tilde{\Delta}_0}
\newcommand{\lnn}{2\alpha_0 \ln{({E_C\over 2\pi T})}}
\newcommand{\nln}{1+2\alpha_0 \ln{({E_C\over 2\pi T})}}
\draft
\title{Mesoscopic quantum transport: Resonant tunneling in the
presence of strong Coulomb interaction}
\author{Herbert Schoeller and Gerd Sch\"on}
\address{
Institut f\"ur theoretische Festk\"orperphysik, Universit\"at
Karlsruhe, 76128 Karlsruhe, Germany}
\date{\today}
\maketitle
\begin{abstract}
Coulomb blockade phenomena and quantum fluctuations are studied in
mesoscopic metallic tunnel junctions with high charging energies. If
the resistance of the barriers is large compared to the quantum
resistance, transport can  be described by sequential tunneling. Here
we study the influence of quantum fluctuations. They are important
when the resistance is small or the temperature very low. A real-time
approach is developed which allows the diagrammatic classification of
``inelastic resonant tunneling'' processes where different electrons
tunnel coherently back and forth between the leads and the metallic
island. With the help of a nonperturbative resummation technique we
evaluate the spectral density which describes the charge excitations
of the system. From it physical quantities of interest like current
and  average charge can be deduced. Our main conclusions are: An
energy renormalization leads to a logarithmic temperature dependence
of the renormalized system parameters. A finite lifetime broadening
can change the classical picture drastically. It gives rise to a
strong flattening of the Coulomb oscillations for low resistances, but
in the Coulomb blockade regime inelastic electron cotunneling
persists. The temperature where these effects are important are
accessible in experiments. 

\end{abstract}

\section{introduction}
Quantum transport through mesoscopic metallic islands
coupled to large reservoirs has been the subject of many theoretical
and experimental investigations in recent years
\cite{Ave-Lik,Gra-Dev,ZPhB}. The
small size of these systems implies a strong Coulomb
interaction which gives rise to a variety of single-electron
phenomena. When the temperature T and the voltage $eV$ are
low compared to the charging
energy $E_C$, tunneling can be suppressed by the Coulomb
blockade. However, when the energy difference $\Delta_0$ between two adjacent
charge states is
comparable to the temperature or the bias voltage, a current
can flow through
the system. As a consequence, the differential conductance shows a peak
structure as function of an external gate voltage $V_g$ (linear
response) or the bias voltage $V$ (nonlinear response). In the absence
of a dissipative environement and provided that the resistance
$R_T$ of a single barrier is much higher than the quantum resistance
$R_K={h\over e^2}$, i.e. $\alpha_0\equiv{1\over 4\pi^2}{R_K\over R_T}\ll
1$, the smearing of these oscillations is dominated by the temperature.
In this classical regime, transport through the system is achieved by
sequences of uncorrelated tunneling processes, which can be described
by lowest order perturbation theory in the coupling between the leads
and the metallic island. The corresponding classical rates can be used
to set up a master equation
\cite{Kul-She,Schoen-Zai,Bee,Ave-Kor-Lik,Mei-Win-Lee}, from which the
charge average and the current can
be calculated. For very low temperatures, however, or when the
coupling $\alpha_0$ becomes larger, quantum fluctuations set in
\cite{Laf-exp,Gla-Mat,Mat,Zai,Gra,Fal-Scho-Zim,Zwe} and the classical picture
breaks down for two reasons.

First, the resummation of the leading logarithmic terms in $\alpha_0
\ln{({E_C\over 2\pi T})}$ leads to a renormalization of the
difference in the charging energies, $\Delta_0$,
and the dimensionless conductance $\alpha_0$.
Provided that $\ds\le
T$, we find
\begin{equation} \label{1}
\tilde{\Delta}_0={\Delta_0\over 1+2\alpha_0 \ln{({E_C\over 2\pi T})}}
\qquad\quad ; \quad\qquad
\tilde{\alpha}_0={\alpha_0 \over 1+2\alpha_0 \ln{({E_C\over 2\pi T})}}\,\,.
\end{equation}
Throughout this work, we set $\hbar=k=1$
and consider only the
case of a wide junction with many transverse channels. The
renormalization is important if $T\le {E_C\over
2\pi}\exp[-{1/(2\alpha_0)}]$, which is an experimentally
relevant temperature if $\alpha_0$ is not too small.

Secondly, coherent processes where the electrons can tunnel an
arbitrary number of times between the leads and the island (resonant
tunneling) give rise to an energy and temperature dependent
broadening of the charge states. Thus, one can overcome the
Coulomb-blockade not only by thermal activations but also by quantum
fluctuations due to higher-order processes. For $\tilde{\Delta}_0 \gg T$ this
phenomenon is known as inelastic cotunneling \cite{Ave-Odi,Ave-Naz}
and can be described in second order
perturbation theory in $\alpha_0$. Important results of the
present work are that higher order tunneling processes describe the
crossover from cotunneling to sequential tunneling and can also give
rise to important corrections to the classical result in the regime
$\tilde{\Delta}_0\le T$
where the conductance reaches its maximum value.
We have called such processes ''inelastic resonant tunneling''
since, for a large number of transverse channels, all the electrons
taking part in a coherent higher order process will be different. This is
in contrast to the usual mechanism of resonant tunneling through a
single level in a quantum dot where only one electron is involved in
any coherent process.
We will show in this work that finite life-time effects due to
resonant tunneling in metallic systems are especially important
when $\pi\tilde{\alpha}_0\sim 1$. For finite
temperatures this can be realized for a sufficiently large value for
$\alpha_0$.

In systems with low tunneling barriers we expect that effects
due to energy renormalization and finite life-times are
simultaneously observable in a real experiment at realistic temperatures.
Both are important since $\pi\as\ll 1$ is equivalent to
$2\alpha_0 \ln({E_C\over2\pi T})\ll 1$ except for very low
temperatures which are experimentally not accessible.

Charge fluctuations in the equilibrium case (e.g. in the single electron box)
have been studied before by many authors. In ref.\cite{Gra} a
systematic perturbation expansion has been performed up to second
order in $\alpha_0$ including all possible charge states. This is a good
approximation if the parameter $2\alpha_0 \ln{({E_C\over 2\pi T})}$
(and consequently also $\pi\tilde{\alpha}_0$) is small compared to unity,
i.e. the energy renormalization and the finite broadening are accounted
for in a perturbative way. In ref.\cite{Mat,Zai,Fal-Scho-Zim} the
leading logarithmic terms together with certain improvements for
larger values of $\alpha_0$ \cite{Fal-Scho-Zim} have been considered in
the two charge state approximation. However finite
broadening effects are neglected. These approaches are valid in the
low temperature regime where $\pi\tilde{\alpha}_0\ll 1$ and lead to
the same renormalization effects as given by Eq.(\ref{1}).
In the nonequilibrium case (SET-transistor), the crossover from
sequential tunneling to inelastic cotunneling has been studied in
\cite{Kor-etal,Naz,Ave,Laf-Est} by introducing a finite and
constant lifetime into the expression of electron cotunneling. The
results are valid in the parameter regime
$2\alpha_0 \ln{({E_C\over 2\pi T})}\ll 1$ and $\pi\as\ll 1$
where renormalization effects and the energy
dependence of the finite life-time can be neglected. In this regime
several experiments have confirmed the theoretical predictions
\cite{Pas-etal,Gee-exp}.

In the present work, we develop a systematic diagrammatic
technique in real time to identify the processes of sequential
tunneling, inelastic cotunneling and resonant tunneling. If we concentrate
on two charge states, we can resum the corresponding diagrams
analytically and obtain closed expressions for the density matrix and
all Green's functions. The spectral density describing the charge
excitations of the system contains an
energy renormalization as well as a finite broadening. Both are retained,
which is crucial for a conserving theory which obeys
sum rules and current conservation. The starting
point of our technique, the real time representation of the density
matrix, is closely related to path integral representations describing
dissipation \cite{Fey-Ver,Wei-Buch} or tunneling in
metallic junctions \cite{Schoen-Zai,Eck-Scho-Amb}. The method allows
us to perform a nonperturbative analysis in the coupling to the
reservoirs while taking into account exactly the strong correlations
due to the Coulomb interaction. Thus, usual Green's
function techniques, either for equilibrium \cite{Fet-Val,Abr-Gor-Dzy} or
nonequilibrium \cite{Ram-Smi,Mah} systems, can not be used
here. The same problem arises in the context of local, strongly
correlated Fermi systems like the Kondo or Anderson model
\cite{Kei-Mor,Bic,Hew}. For these systems diagrammatic
techniques very similiar to our ones have been used by Barnes
\cite{Bar} for the
equilibrium case starting from a slave-boson description. Another
example is a work of Rammer \cite{Ram} who developed the same
graphical language within a density-matrix description of the dynamics
of a particle coupled to a heat bath. Finally, the technique
presented in this work can also be formulated very elegantly in terms
of Liouville operators using projection operator techniques developed
in Ref.\cite{Los-Schoeller}. These relationships
as well as the generalization to other systems like
quantum dots with arbitrary many-particle correlations,
time-dependent Hamiltonians and influences of the electrodynamic
environement will be the subject of forthcoming works.\\

\section{Hamiltonian and physical quantities}
We consider a small metallic island coupled via tunneling barriers to two
leads. In addition it is coupled capacitatively to an external gate
voltage (SET-transistor, see Fig.(\ref{fig1})). This system is
described by the following Hamiltonian
\begin{equation}\label{2}
H=H_L+H_R+H_I+V+H_T=H_0+H_T.
\end{equation}
Here
\begin{equation}\label{3}
H_r=\sum_{kn}\epsilon^r_{kn}a^\dagger_{krn}a_{krn}
\qquad,\qquad
H_I=\sum_{ln}\epsilon_{ln} c^\dagger_{ln} c_{ln}
\end{equation}
describe the noninteracting electrons in the two leads $r=L,R$ and on
the island where n is the transverse channel index which includes the
spin. The wave vectors k and l numerate the states of the
electrons for fixed r and n (a subindex
$k_{rn}$ or $l_n$ has been omitted for simplicity).
The Coulomb interaction V obtained by straightforward
electrostatic considerations \cite{Ave-Lik,Kul-She,Gla-Mat} is given by
\begin{equation}\label{4a}
V=E_C({\hat{Q}_I\over e}-{q_x\over e})^2
\end{equation}
where $E_C={e^2\over 2C}$ is the charging energy, $q_x=C_L V_L+C_R V_R+
C_g V_g$, $C=C_L+C_R+C_g$ and $V_s,C_s$ (s=L,R,g) are the voltages
and capacitances of the circuit according to Fig.(\ref{fig1}). $\hat{Q}_I$
denotes the charge operator of the island and is given by ($e<0$)
\begin{equation}\label{4b}
\hat{Q}_I=e(\hat{N}_I-N_+)
\end{equation}
where $\hat{N}_I=\sum_{ln}c^\dagger_{ln}c_{ln}$ is the particle number
operator and $-eN_+$ the positive background charge of the island.

The charge transfer processes due to tunneling are described
by
\begin{equation}\label{5a}
H_T=\sum_{r=L,R}\sum_{kln}(T^{rn}_{kl}a^\dagger_{krn}c_{ln}+h.c.)
\end{equation}
where $T^{rn}_{kl}$ are the tunneling matrix elements.

For convenience, one can also introduce an auxiliary set of discrete
charge states $|N\rangle$ with $N=0,\pm 1,\pm 2,\ldots$
and write the Coulomb interaction and the tunneling part in the form
(see e.g. \cite{Mat,Gra,Gui-Schoen,Ing-Naz})
\begin{equation}\label{4}
V(\hat{N})=E_C(\hat{N}-{q_x\over e})^2
\end{equation}
\begin{equation}\label{5}
H_T=\sum_{r=L,R}\sum_{kln}(T^{rn}_{kl}\,a^\dagger_{krn}c_{ln}
e^{-i\hat{\Phi}}\,+\,h.c.)
\end{equation}
provided that we impose the constraint $eN=Q_I$ in calculating the
trace of any physical quantity. Thus $eN$ can be interpreted as the
charge on the island which is now separated from all the other degrees
of freedom although they are of course not independent due to the constraint.
$\hat{\Phi}$ is the phase operator canonical conjugate to $\hat{N}$,
i.e. $[\hat{\Phi},\hat{N}]=i$ and $e^{\pm i\hat{\Phi}}$
changes the charge by $\pm 1$
(note that N can
take all integer numbers from $-\infty$ to $\infty$ here, so that the phase
operator is  well defined and the charge transfer operator $e^{\pm
i\hat{\Phi}}$ is unitary).

Since both representations are equivalent, it is a matter of taste
which one is used. However, for the metallic case the energy spectrum
of the island is dense and the total particle number $N_I$ on the
island is very large. In this case the
constraint $eN=Q_I$ can be disregarded since any canonical average
over island operators with fixed $N_I$ (or equivalently fixed $N$ due
to the constraint) can be replaced by the corresponding grandcanonical
average. Such an approximation is of course not valid if we consider
the case of a discrete level spectrum in a small island like a quantum
dot. In this case, the particle number on the dot can be very low and
one has to use the representation (\ref{4a}) and (\ref{5a}).

Without the constraint, the separation of the charge degree of freedom
is very convenient and effectively
has also been used in other treatments like
the path integral formalism \cite{Eck-Scho-Amb}. The
technical advantage is that the lead and island electron field
operators occur now bilinear in the
part $H_0$ of the Hamiltonian (\ref{2}) since the Coulomb interaction
(\ref{4}) contains only the charge degree of freedom. In section 3 we
will see that this allows for a straighforward application of Wick's
theorem to integrate out all degrees of freedom of the leads and the island.

The two leads as well as the island are
treated as large equilibrium reservoirs which are not affected
significantly by coherent tunneling processes where only a small number of
electrons are involved. Thus we describe the electrons in these
three ''reservoirs'' initially by Fermi
distribution functions $f_i(E)=1/(\exp(\beta(E-\mu_i)+1)$ (i=L,R,I)
with fixed chemical potentials $\mu_i$ or equivalently by the
grandcanonical density matrix
\begin{equation}\label{6}
\rho_{res}^0={1\over
Z_{res}^0}\\exp\left[\beta\!\!\sum_{i=L,R,I}(H_i-\mu_i\hat{N_i})
\right]
\end{equation}
where $\beta=1/T$ and $\hat{N}_i$ are the number of electrons in the
leads and the island, respectively.
In the following, we will always set the chemical potential
$\mu_I$ of the island to zero.

Furthermore, we assume that at the initial time $t_0$ there are no
correlations between the reservoirs and the discrete charge
states. This means that the initial density matrix of the total system
reads $\rho_0=\rho_{res}^0 \hat{P}^0$ where
$\hat{P}^0$ is a diagonal operator
with matrix elements $P_N^0=\langle N|\hat{P}^0|N\rangle$ and
$P_N^0=P_N(t_0)$ is an arbitrary initial
distribution for the discrete charge states at time $t_0$.

After switching on the tunneling part $H_T$ adiabatically, the
probability $P_N(t)$ for a certain charge N on the island at time t
can be calculated from
\begin{equation}\label{7}
P_N(t)=\langle N| e^{-iH(t-t_0)}\,\rho_0\,
e^{iH(t-t_0)}|N \rangle .
\end{equation}
The stationary distribution follows from
\begin{equation}\label{8}
P_N^{st}=\lim_{t\rightarrow\infty}P_N(t)=\lim_{t_0\rightarrow -\infty}
P_N(0)
\end{equation}
and will turn out to be independent of the initial choice for
$P_N^0$. Note that $P_N^{st}$ is not an equilibrium distribution if the
chemical potentials $\mu_{L/R}$ of the leads are different.

The average charge of the island, which is an experimentally measurable
quantity via the electrostatic potential of the island, can be
calculated from
\begin{equation}\label{9}
\bar{N}=\sum_N N\,P_N^{st}.
\end{equation}

The current flowing through the barriers $r=L/R$ is defined by
$I_r(t)=e{d\over dt}{\langle\hat{N}_r(t)\rangle}_{\rho_0}$. After a
straightforward calculation one obtains
\begin{equation}\label{10}
I_r(t)=2eIm\left\{\sum_{kln}T^{rn}_{kl}{\langle(a^\dagger_{krn}
c_{ln}e^{-i\hat{\phi}})(t)\rangle}_{\rho_0}\right\}
\end{equation}
and for the stationary current
\begin{equation}\label{11}
I^{st}_r=\lim_{t\rightarrow\infty}I_r(t)=\lim_{t_0\rightarrow -\infty}
I_r(0).
\end{equation}

At the end of section 4 (see (\ref{46c})) we will show that the stationary
current can also be related to the real-time correlation
functions
\begin{eqnarray}
C^>(t,t^\prime)&=&-i{\langle
e^{-i\hat{\phi}(t)}e^{i\hat{\phi}(t^\prime)}
\rangle}_{\rho_0}\label{12}\\
C^<(t,t^\prime)&=&i{\langle
e^{i\hat{\phi}(t^\prime)}e^{-i\hat{\phi}(t)}
\rangle}_{\rho_0}\label{13}.
\end{eqnarray}
Both are independent quantities since we do not assume
equilibrium here. In the stationary limit, i.e. $t_0\rightarrow -\infty$,
the correlation functions depend only on the relative time difference
$\tau=t-t^\prime$ and we can define the Fourier transform
\begin{equation}\label{14}
C^>(\omega)=\int\,d\tau\,e^{i\omega\tau}C^>(\tau)
\end{equation}
and analog for $C^<(\omega)$.

 From a theoretical point of view, an interesting quantity is also the
spectral density
\begin{equation}\label{15}
A(\omega)={1\over 2\pi i}[C^<(\omega)-C^>(\omega)]
\end{equation}
which describes the charge excitations of the system. At the end of
section 4 we will see that within our approximations all quantities of
interest like the probability $P_N^{st}$, the current $I_{L/R}^{st}$ and the
correlation functions $C^>,C^<$  can be related to the spectral
density. Thus, its specific form, which is related to energy
renormalization and finite life-time broadening effects, will be the
central point of our analysis.\\

\section{Diagrammatic technique}
To start with, we consider the probability distribution (\ref{7}) and
write it in the form
\begin{equation}\label{16}
P_N(t)=\sum_{N^\prime}P_{N^\prime}^0 \,Tr_{res}\,\rho_{res}^0\,\langle
N^\prime|\, T^{(+)}
e^{i\int_{t_0}^t d\tau H_T(\tau)_0}\,|N\rangle\,\langle N|\,T^{(-)}
e^{-i\int_{t_0}^t d\tau H_T(\tau)_0}\,|N^\prime\rangle
\end{equation}
where $Tr_{res}$ is the trace over all reservoirs, $T^{(+)}$ ($T^{(-)}$) are
the time-ordering (anti-time-ordering) operators and $H_T(\tau)_0$
denotes the tunneling part of the Hamiltonian (\ref{2}) in the interaction
representation with respect to $H_0$. Note that the Coulomb
interaction V is included here in $H_0$ and will be treated
exactly in the following. Eq.(\ref{16}) has also been used as a
starting point by other
authors, e.g. by Feynman and Vernon \cite{Fey-Ver} or Caldeira and
Leggett \cite{Cal-Leg} (see also \cite{Wei-Buch}) for a system coupled to
a heat bath or by Eckern et al \cite{Eck-Scho-Amb}
for a tunnel junction. After integrating out the
reservoirs within a real-time path integral representation
these authors obtained an effective action where the two propagators
occuring in Eq.(\ref{16}) are coupled to each other. Expanding the
exponential of
this action in terms of the tunneling, one can arrive at a graphical
language in real-time space \cite{Schoen-Zai}.

In this work we will use a different approach which is
easily generalized to other systems such as quantum dots,
Anderson and Kondo models, etc. \cite{Schoeller}. The procedure is first
to expand the propagators in $H_T$
\begin{equation}\label{17}
T^{(\pm)}e^{\pm i\int^t_{t_0}d\tau H_T(\tau)_0}=\sum_{m=0}^\infty (\pm i)^m
\int^t_{t_0}d\tau_1\int^{\tau_1}_{t_0}d\tau_2\ldots\int^{\tau_{n-1}}
_{t_0}d\tau_n \,\,T^{(\pm)}\{H_T(\tau_1)\cdots H_T(\tau_n)\}
\end{equation}
and, in a second step, insert the form (\ref{5}) for $H_T$ and apply
Wick's theorem with respect to the reservoir field operators
$a^{(\dagger)}_{krn}$ and $c^{(\dagger)}_{ln}$. This is possible since
$H_0$ is bilinear in these operators. As a consequence, the
vertices corresponding to the charge transfer operators $e^{\pm
i\hat{\phi}}$ become coupled by reservoir lines, either from the two
leads or the island. This is indicated in a graphical language by
solid lines for the leads and wiggly lines for the island as shown in
Fig.\ref{fig2}. There the upper (lower) horizontal line corresponds to the
forward (backward) propagator and the vertices change the discrete
charge states as indicated by the particle numbers associated with the
horizontal lines. The two charge states at the left end of the diagram
at time $t_0$ are the same and correspond to
the initial probability distribution $P^0_{N^\prime}$ (if nothing is
written there in a certain graph we imply this automatically).
The two charge states at the right end at time t are chosen equal to N if
we want to calculate $P_N(t)$. Graphically we indicate the
relationship of the forward and backward propagator at the right end point
symbolically by $|N\rangle\langle N|$.
Furthermore we assign to each line a certain energy,
the Coulomb energy $V(N)$ to the horizontal lines and the energies
$\epsilon$ ($E$) to the lead (island) lines.
The rules to translate a certain diagram in time space are the
following ones:
\begin{description}
\item[ 1.   ]
Assign a factor $e^{-i x_j \tau_j}$ to each vertex where $\tau_j$ is
the time variable and $x_j$ is the difference of the energies entering
the vertex minus all energies leaving the vertex.
\item[ 2.   ]
To each loop formed by $2s$ reservoir lines we assign a factor
\begin{eqnarray}
\lefteqn{\alpha^{\sigma_1\ldots\sigma_s,\eta_1\ldots\eta_s}_
{r_1\ldots r_s}(\epsilon_1\ldots\epsilon_s,E_1\ldots E_s)=}
\hspace{2cm}\nonumber\\&&\nonumber\\&&
=\sum_{l_1\ldots l_s}\Gamma^{r_1n\sigma_1\eta_1}_
{l_1 l_2}(\epsilon_1,E_1)\Gamma^{r_2n\sigma_2\eta_2}_{l_2
l_3}(\epsilon_2 ,E_2)\cdots
\Gamma^{r_sn\sigma_s\eta_s}_{l_s l_1}(\epsilon_s ,E_s)\label{18}
\end{eqnarray}
where $\epsilon_1,\ldots,\epsilon_s$ and $E_1,\ldots,E_s$ are the
energies associated to the lead and island electrons, respectively,
ordered in the direction of the loop, $r_j=L/R$ specifies the left or
the right lead and each sign $\sigma_j,\eta_j=\pm$ indicates whether the
reservoir line with energy $\epsilon_j,E_j$
runs to a lower (+) or a larger (-) time with respect to the
closed time path formed by the two propagators.
The quantities $\Gamma$ are defined as
\begin{equation}\label{19}
\Gamma^{rn\sigma\eta}_{ll^\prime}(\epsilon,E)=\sum_k T^{rn}_{kl}
{T^{rn}_{kl^\prime}}^*
\delta(\epsilon-\epsilon^r_{kn})\delta(E-\epsilon_{ln})
f^{\sigma}_r(\epsilon)f^{\eta}_I(E)
\end{equation}
where $f_i^+=f_i$ (i=L,R,I) is the Fermi distribution function and
$f_i^-=1-f_i$.
\item[  3.   ]
The prefactor is given by $(-i)^M(-1)^m(-1)^c$, where M is the total
number of internal vertices, m the number of internal vertices on the
backward propagator and c the number of crossings of reservoir lines.
\end{description}
Finally one has to integrate over all time variables $\tau_j$ from
$t_0$ to t and over all energy variables of the reservoir lines.

For the current (\ref{10}) and the correlation functions (\ref{12},\ref{13}),
the procedure is completely analog, the only difference is that
additional external vertices have to be added. As shown in
Fig.\ref{fig3} one has to introduce an external
vertex with two reservoir lines at the right end of the diagram to
calculate the current, and two external vertices to calculate the correlation
functions.

So far, the diagrammatic rules are formulated in time space
which is not the most convenient one to calculate stationary transport
properties. Therefore, using (\ref{8}) and (\ref{11}) we set $t=0$ and
$t_0=-\infty$ in
every diagram for $P_N(t)$ and $I_r(t)$ and evaluate the time
integrals analytically
for a given fixed ordering of all time variables by using the identity
\begin{eqnarray}\lefteqn{\int^0_{-\infty}d\tau_1\int^0_{\tau_1}d\tau_2
\ldots\int^0_{\tau_{M-1}}d\tau_M\,e^{-ix_1\tau_1}e^{-ix_2\tau_2}\cdots
e^{-ix_M\tau_M}e^{\eta\tau_1}\,=}\hspace{2cm}\nonumber\\&&
=\,i^M{1\over x_1+i\eta}\,\cdot\,{1\over x_1+x_2+i\eta}\cdots{1\over
x_1+x_2+\ldots +x_M+i\eta}\label{20}
\end{eqnarray}
where $e^{\eta\tau_1}$ ($\eta\rightarrow 0^+$) is a convergence factor
which is related to the adiabatic turn on of the coupling part $H_T$. The
rules in energy space read
\begin{description}
\item[  1.   ]
For each auxiliary vertical line which does not cut any vertex and has
always a vertex to its left, we
assign a resolvent ${1\over \Delta E +i\eta}$ where $\Delta E$ is the
difference of all energies crossing the vertical line from right to left minus
all energies crossing it from left to right (see Fig.\ref{fig4}).
\item[  2.   ]
For each loop we assign the same factor as given by Eq.(\ref{18}).
\item[  3.   ]
The prefactor is given by $(-1)^m (-1)^c$, where m is the total number
of internal vertices on the backward propagator and c the number of
crossings of reservoir lines.
Furthermore, if we are calculating the current, we have to multiply
with an additional factor $-ie$ and we have to assign a factor $-1$ if
the external vertex to the right has an incoming lead line.
\end{description}
For the Fourier transform (\ref{14}) of the correlation functions we
write
\begin{equation}\label{21}
C^>(\omega)=\int^0_{-\infty}d\tau\,\left[e^{-i\omega\tau}\lim_{t_0
\rightarrow -\infty}C^>(0,\tau)+e^{i\omega\tau}\lim_{t_0\rightarrow
-\infty}C^>(\tau,0)\right]
\end{equation}
and again calculate all time integrals analytically by using (\ref{20})
and fixing the time ordering of all time variables including the
integration variable $\tau$. Rule 1 has then to be
supplemented by
\begin{description}
\item[  4.   ]
If an auxiliary vertical line lies between the two external vertices,
we have to add $\pm\omega$ to the energy difference $\Delta E$ when
the vertex $e^{\pm i\hat{\phi}}$ lies to the left of the auxiliary
line (see Fig.\ref{fig5}). If we imagine a virtual line which
connects the two external vertices from $e^{-i\hat{\phi}}$ to
$e^{i\hat{\phi}}$ and assign an energy $\omega$ to it, we can simulate
the same by consequently applying rule 1 including this virtual line.
\end{description}
The graphical representation of $C^>(\omega)$ and $C^<(\omega)$ in
energy space is given in Fig.\ref{fig6}.\\

\section{Physical processes}
Usually, tunneling processes within the SET-transistor are described
by a classical master equation
\cite{Ave-Lik,Schoen-Zai,Bee,Ave-Kor-Lik,Mei-Win-Lee,Gei-Scho}
\begin{equation}\label{22}
\dot{P}^{(0)}_N(t)=\sum_{N^\prime\neq N}\left[P_{N^\prime}^{(0)}(t)
\gamma_{N^\prime N}-P^{(0)}_N(t)\gamma_{NN^\prime}\right]
\end{equation}
where $\gamma_{N^\prime N}$ are classical transition rates
calculated by the golden rule in second order perturbation theory in
$H_T$
\begin{equation}\label{23}
\gamma_{N^\prime N}=2\pi\sum_r\left[
\alpha^-_r(\Delta_N)\delta_{N^\prime,N+1}
+\alpha^+_r(\Delta_{N-1})\delta_{N^\prime,N-1}\right].
\end{equation}
Here $\Delta_N=V(N+1)-V(N)$ is the change of the Coulomb energy if we
increase the excess particle number by one and
\begin{equation}\label{24}
\alpha^{\pm}_r(\omega)=\int dE\,\alpha^{\pm,\mp}_r(\omega+E,E)
\end{equation}
are the transition rates for tunneling in and out processes. Assuming
constant tunneling matrix elements $T^{rn}=T^{rn}_{kl}$ and neglecting
the energy dependences of the density of states $\rho_r^n(\epsilon),
\rho^n_I(E)$ in the leads and the island, we obtain the
well-known expressions
\begin{equation}\label{25}
\alpha^{\pm}_r(\omega)=(\omega-\mu_r)\,\alpha^r_0\,n^{\pm}_r(\omega),
\end{equation}
where
\begin{equation}\label{25a}
\alpha^r_0=\sum_n\,|T^{rn}|^2\rho^n_r\rho^n_I={1\over 4\pi^2}
{R_K\over R_T^r}
\end{equation}
is the ratio of the quantum resistance $R_K={h\over e^2}$ to the
tunneling resistance $R^r_T$ of barrier r,
$n^+_r(\omega)=1/(\exp(\beta(\omega-\mu_r))-1)$ is the Bose
function and $n^-_r=1+n_r^+$. Furthermore, we
introduce the notations
$\alpha_0=\sum_r\alpha_0^r$ , $\alpha^\pm(\omega)=\sum_r\alpha^\pm_r
(\omega)$ , $\alpha_r(\omega)=\alpha^+_r(\omega)+\alpha^-_r(\omega)$
and $\alpha(\omega)=\alpha^+(\omega)+\alpha^-(\omega)$.

Physically, the master equation (\ref{22}) describes sequences of
uncorrelated lowest order processes (sequential tunneling) where each
single process describes one electron entering or leaving the
island. Retardation effects and higher-order correlated tunneling
processes are neglected within this approach. We can write (\ref{22})
in the form
\begin{equation}\label{26}
\dot{P}_N^{(0)}(t)=\sum_{N^\prime}P^{(0)}_{N^\prime}(t)
\bar{\gamma}_{N^\prime N}
\end{equation}
with $\bar{\gamma}_{N^\prime N}=\gamma_{N^\prime
N}-\delta_{N^\prime N}\sum_{N^{\prime\prime}}\gamma_{NN^{\prime
\prime}}$. We will now show that the exact kinetic
equation has a similiar form
\begin{equation}\label{27}
\dot{P}_N(t)=\sum_{N^\prime}\int^t_{t_0}dt^\prime\,P_{N^\prime}(t^\prime)
\tilde{\Sigma}_{N^\prime N}(t^\prime-t).
\end{equation}
$\tilde{\Sigma}$ denotes the sum of all possible correlated
processes. Now non-Markovian effects are included. For the stationary
solution $P_N^{st}$ this gives
\begin{equation}\label{28}
0=\sum_{N^\prime}P^{st}_{N^\prime}\Sigma_{N^\prime N}\,\,,
\end{equation}
where $\Sigma_{N^\prime
N}=i\int^0_{-\infty}d\tau\tilde{\Sigma}_{N^\prime N}(\tau)$.
This equation can immediately be obtained by using our diagrammatic
technique in energy space. The self-consistent equation for
$P^{st}_N$ is shown in Fig.\ref{fig7}. Here $\Sigma$ is defined
graphically as the sum of all possible irreducible self-energy diagrams, which
are defined such that any vertical line cutting them will cross some
reservoir line (see Fig.\ref{fig8}). According to our diagrammatic
rules 1,2 and 3 in energy space, the equation corresponding to
Fig.\ref{fig7} reads
\begin{equation}\label{29}
P_N^{st}=P^0_N+\sum_{N^\prime}P_{N^\prime}^{st}\Sigma_{N^\prime N}
{1\over i\eta}.
\end{equation}
Multiplying with $i\eta$ and performing the limit $\eta\rightarrow
0^+$, we arrive at (\ref{28}) and can identify the sum of all possible
correlated tunneling processes diagrammatically by the irreducible
self-energy $\Sigma$. Furthermore, the solution of Eq.(\ref{28}) is independent
of the initial choice for $P_N^0$ which has dropped out of Eq.(\ref{29})
in the limit $\eta\rightarrow 0^+$. In the same way, one
can also set up a self-consistent equation in time space to get the
time-dependent kinetic equation (\ref{27}).

By moving the last vertex to the right of each diagram of $\Sigma$ up
or down, one can easily show that
\begin{equation}\label{30}
\sum_{N^\prime}\Sigma_{NN^\prime}=0
\end{equation}
which provides the possibility to write Eq.(\ref{28}) also in the form
\begin{equation}\label{31}
0=\sum_{N^\prime\neq N}\left(P^{st}_{N^\prime}\Sigma_{N^\prime N}-
P^{st}_N\Sigma_{NN^\prime}\right).
\end{equation}
This has the form of a balance equation for each charge state
$|N\rangle$. Furthermore, by changing the vertical position of all
vertices and the direction of all reservoir lines one can easily see
that $\Sigma_{N^\prime N}$ is purely imaginary, i.e. there
exists a real solution of (\ref{28}) and (\ref{31}).

Calculating $\Sigma_{N^\prime N}$ in first order in $\alpha_0$ (see
Fig.\ref{fig9}), we get
\begin{equation}\label{32}
\Sigma^{(1)}_{N+1,N}=2\pi i\alpha^-(\Delta_N)\,\,,\,\,
\Sigma^{(1)}_{N-1,N}=2\pi i\alpha^+(\Delta_{N-1})\,,
\end{equation}
which, after insertion in Eq.(\ref{28}), leads to the classical result
(\ref{26}) in the stationary limit.

Before starting to include higher-order processes in $\Sigma$, we will
introduce two important approximation which will simplify the
following analysis considerably.

First we assume that the number Z of transverse channels is very
large. Since each loop contribution (\ref{18}) is proportional to
$Z\Gamma^s$, this has the consequence that all loops with $s=1$ will
dominate in each given perturbation order in $\Gamma$. Thus, we can
restrict ourselves to the loops with two reservoir
lines which give the contribution $\alpha^{\pm,\mp}_r(\epsilon,E)$.
Since the rest of the diagram depends only on $\omega=\epsilon-E$, we
can integrate $\alpha^{\pm,\mp}_r(\omega+E,E)$ over E and get the
contribution $\alpha^{\pm}_r(\omega)$ defined by
(\ref{24},\ref{25}). Furthermore, we represent all loops with two
reservoir lines
from now on graphically by single solid lines with one energy variable
$\omega$ (see Fig.\ref{fig10}).

Secondly, if the charging energy $E_C$ is large compared to the energy
difference of the two charge states $\Delta_0=V(1)-V(0)$ and the bias
voltage $V=V_L-V_R$, we can use the two charge state
approximation. This means that we consider only the charge states
$N=0,1$.

The classical approach breaks down for low temperatures or for large
values for the dimensionless conductance $\alpha_0$. Specifically we
will see in section 5 that the classical master equation is valid for
$\alpha_0 \ln{({E_C\over 2\pi T})}\ll 1$. To go beyond this regime we
consider now higher-order correlated tunneling processes in
$\Sigma$. Fig.\ref{fig11} shows an example in second order in
$\alpha_0$ which represents a part of the rate of inelastic electron
cotunneling \cite{Ave-Odi,Ave-Naz} when the leads $r$ and $r^\prime$ are
different. Here one electron enters the island from lead $r$, the
system stays in a virtual intermediate charge state with $N=1$ and
finally another electron leaves the island via lead
$r^\prime$. However, the calculation of all second order diagrams for
$\Sigma_{01}=-\Sigma_{00}$ or $\Sigma_{10}=-\Sigma_{11}$ is plagued by
several irregular integrals which occur since the complete self-energy
depends nonanalytically on $\alpha_0$. Thus, we will directly perform a
nonperturbative resummation of higher-order diagrams and discuss the
second order result as a limiting case at the end. By this, we are
also able to clarify certain singularities which are present in the
usual expressions for inelastic electron cotunneling
\cite{Ave-Odi,Ave-Naz}. Furthermore, we are able to go beyond the
cotunneling theory and can investigate the influence of resonant tunneling
processes which will modify the classical result of sequential
tunneling significantly if $\alpha_0 \ln{({E_C\over 2\pi T})}\sim 1$.

A very illustrative example of a resonant tunneling process is shown
in Fig.\ref{fig12}. Here the charge of the island is alternating
between 0 and 1 via an infinite number of intermediate virtual
states. The electrons going back and forth between the leads and the
island or tunneling from one lead to another via the island
are all different since we have assumed a large number of
transverse channels. This is in contrast to the usual mechanism of
resonant tunneling where only one level of the island is involved. Of
course, the diagram of Fig.\ref{fig12} is not the only one which is
important to describe higher-order processes. In order to get a
systematic criterion which diagrams might be relevant ones,
we take a look at the states corresponding to the upper and lower
horizontal lines before integrating out the reservoir degrees of
freedom. Two such states which lie on the same vertical cut define a
matrix element of the total density matrix (i.e. reservoirs plus
charge states). Our criterion now is that
we take into account only those nondiagonal matrix elements of the
total density matrix, which differ at most by
one electron-hole pair excitation in the leads or (equivalently) in
the island. Graphically this means that any vertical line will at most
cut through two solid lines each representing a pair of one lead and
one island line (see Fig.\ref{fig13} for examples). The sum of all these
diagrams can be represented as shown in
Figs.\ref{fig14}-\ref{fig16}. We get for $N=0,1$
\begin{equation}\label{33}
\Sigma_{N1}=-2i\,Im\int d\omega\sum_r\phi^r_N(\omega)
\end{equation}
together with the self-consistent equation
\begin{eqnarray}
\lefteqn{\phi^r_N(\omega)=\Pi(\omega)\left[\alpha^+_r(\omega)\delta_{N0}
-\alpha^-_r(\omega)\delta_{N1}-\right.}\hspace{2cm}\nonumber\\&&
\left.-\,\alpha_r(\omega)\int d\omega^\prime\,{1\over
\omega-\omega^\prime+i\eta}
\sum_{r^\prime}\phi^{r^\prime}_N(\omega^\prime)^*\right],\label{34}
\end{eqnarray}
where $\Pi(\omega)=1/[\omega-\Delta_0-\sigma(\omega)]$ and
\begin{equation}\label{35}
\sigma(\omega)=\int d\omega^\prime{\alpha(\omega^\prime)\over
\omega-\omega^\prime+i\eta}.
\end{equation}
is a self-energy which will play an important role in the following.
Defining
\begin{eqnarray}
\psi_+(\omega)&=&{\phi^r_0(\omega)\over \alpha_r(\omega)}-
{\alpha^+_r(\omega)\over\alpha_r(\omega)}\Pi(\omega)+{\alpha^+(\omega)
\over \alpha(\omega)}\Pi(\omega)\label{36a}\\
\psi_-(\omega)&=&{\phi^r_1(\omega)\over \alpha_r(\omega)}+
{\alpha^-_r(\omega)\over\alpha_r(\omega)}\Pi(\omega)-{\alpha^-(\omega)
\over \alpha(\omega)}\Pi(\omega)\label{36b}
\end{eqnarray}
which are quantities independent of r due to Eq.(\ref{34}), we obtain
the integral equation
\begin{equation}\label{37}
[\omega-\Delta_0-\sigma(\omega)]\,\psi_{\pm}(\omega)=\pm{\alpha^\pm(\omega)
\over \alpha(\omega)}-\int d\omega^\prime\,{\alpha(\omega^\prime)\over
\omega-\omega^\prime+i\eta}\psi_\pm(\omega^\prime)^*.
\end{equation}
Since $Im\,\sigma(\omega)=-\pi\alpha(\omega)$, the solution of this
integral equation is
\begin{equation}\label{38}
Im\,\psi_\pm(\omega)=\mp\pi{\lambda_\pm\over\lambda}|\Pi(\omega)|^2
\end{equation}
where
\begin{equation}\label{39}
\lambda_\pm=\int d\omega\alpha^\pm(\omega)|\Pi(\omega)|^2\qquad,\qquad
\lambda=\int d\omega |\Pi(\omega)|^2
\end{equation}
and the real part of $\psi_\pm(\omega)$ can be obtained from
(\ref{37}) and the
Kramers-Kronig relation although it is not necessary for the
following. From Eq.(\ref{38}) we can calculate $Im\,\phi^r_N(\omega)$ with
the help of (\ref{36a},\ref{36b}) and obtain from Eq.(\ref{33}) a
nonperturbative expression for the transition matrix elements
\begin{eqnarray}
\Sigma_{01}&=&-\Sigma_{00}=2\pi i{\lambda_+\over\lambda}\label{40a}\\
\Sigma_{10}&=&-\Sigma_{11}=2\pi i{\lambda_-\over\lambda}\label{40b}.
\end{eqnarray}
Inserting these quantities in the kinetic equation (\ref{31}) and
solving it we obtain for the stationary probabilities
\begin{equation}\label{41}
P_0^{st}=\lambda_-\,\,\,,\,\,\,P_1^{st}=\lambda_+\,,
\end{equation}
where $\lambda_+ +\lambda_- =1$ has been used which ensures the
normalization $P_0^{st}+P_1^{st}=1$. Furthermore, both probabilities
are strictly positive. The solution (\ref{41}) is the final result for
the density matrix and will be discussed in detail in section 5.2.

In order to calculate the stationary current $I^{st}_r$ we relate it to
the correlation functions $C^>(\omega)$ and $C^<(\omega)$ as indicated
in Fig.\ref{17} which gives
\begin{equation}\label{42}
I^{st}_r=-ie\int d\omega\left\{\alpha^+_r(\omega)C^>(\omega)+
\alpha^-_r(\omega)C^<(\omega)\right\}.
\end{equation}
This relation is exact in the limit of a very large number of
transverse channels. Otherwise the current will also depend on
correlation functions involving more than two charge transfer
operators $e^{\pm i\hat{\phi}}$.

The correlation functions can now be calculated from the diagrams shown in
Fig.\ref{fig18} where we have used the same criterion as for the
calculation of the density matrix with one exception. If a vertical
line lies between the external vertices we allow for a cut through at
most one solid line. Thereby we have used the fact that the external
vertices also create one electron-hole pair excitation in the system.
Formally we can say that a vertical line between the external vertices
will in addition always cut the virtual line connecting
the external vertices. The sum of all these diagrams gives
\begin{eqnarray}
C^>(\omega)&=&2iIm\left\{P_0^{st}\,\Pi(\omega)-\sum_{N=0,1}\,\sum_{r=L/R}
P_N^{st}\int d\omega^\prime{\phi^r_N(\omega^\prime)^*\over
\omega-\omega^\prime+i\eta}\Pi(\omega)\right\}\label{43a}\\
C^<(\omega)&=&-2iIm\left\{-P_1^{st}\Pi(\omega)^*+\sum_{N=0,1}\,\sum_{r=L/R}
P_N^{st}\int d\omega^\prime{\phi^r_N(\omega^\prime)\over
\omega^\prime-\omega+i\eta}\Pi(\omega)^*\right\}\label{43b}
\end{eqnarray}
which yields with Eqs.(\ref{36a},\ref{36b}),(\ref{38}) and (\ref{41})
\begin{eqnarray}
C^>(\omega)&=&-2\pi i \,\alpha^-(\omega)\,|\Pi(\omega)|^2\label{44a}\\
C^<(\omega)&=&2\pi i \,\alpha^+(\omega)\,|\Pi(\omega)|^2.\label{44b}
\end{eqnarray}
Furthermore, we obtain for the spectral density (\ref{15})
\begin{equation}\label{45}
A(\omega)=\alpha(\omega)\,|\Pi(\omega)|^2
\end{equation}
which is normalized to unity $\int d\omega A(\omega) =1$.

In summary we can
express all our final results in terms of the spectral density
\begin{eqnarray}
P^{st}_0&=&\int d\omega {\sum_r\alpha_r(\omega)f^-_r(\omega)\over
\alpha(\omega)}A(\omega)\label{46a}\\
P^{st}_1&=&\int d\omega {\sum_r\alpha_r(\omega)f^+_r(\omega)\over
\alpha(\omega)}A(\omega)\label{46b}\\
I^{st}_r&=&{e\over h}4\pi^2\int
d\omega\sum_{r^\prime}{\alpha_{r^\prime}
(\omega)\alpha_r(\omega)\over\alpha(\omega)}A(\omega)
[f^+_{r^\prime}(\omega)-f^+_r(\omega)]\label{46c}\\
C^>(\omega)&=&-2\pi i{\sum_r\alpha_r(\omega)f^-_r(\omega)\over
\alpha(\omega)}A(\omega)\label{46d}\\
C^<(\omega)&=&2\pi i{\sum_r\alpha_r(\omega)f^+_r(\omega)\over
\alpha(\omega)}A(\omega)\label{46e}
\end{eqnarray}
where we have used the relation $\alpha^\pm_r(\omega)=\alpha_r(\omega)
f_r^\pm(\omega)$.

These results satisfy conservation laws and sum
rules. Current is conserved, i.e. $\sum_r I^{st}_r=0$, and is zero in
the equilibrium case when $\mu_r=0$. All probabilities are positive,
the spectral
density is normalized to unity and we get the correct relationships
between the correlation functions and the spectral density in the
equilibrium case. The classical result can be recovered from
(\ref{46a}-\ref{46e})
if we use the lowest order approximation in $\alpha_0$ for the
spectral density
\begin{equation}\label{47}
A^{(0)}(\omega)=\delta(\omega-\Delta_0).
\end{equation}
We conclude this section by the observation that quantum
fluctuation effects due to energy renormalization and broadening
manifest themselves in the spectral density via  the real and
imaginary part of the self-energy $\sigma(\omega)$ given in
Eq.(\ref{35}). This quantity will be the main subject of the detailed
discussion in the next section.\\

\section{Results and applications}
\noindent
{\bf\underline{5.1 The spectral density}}\\ \\
Using Eq.(\ref{25}) we can evaluate the self-energy $\sigma(\omega)$
given by Eq.(\ref{35}) as
\begin{equation}\label{48}
\sigma(\omega)=-\sum_r\alpha^r_0\left[R(\omega-\mu_r)+iI(\omega-\mu_r)\right]
\end{equation}
where
\begin{eqnarray}
R(\omega)&=&D(\omega)\omega\left[\psi({E_C\over 2\pi
T})+\psi(1+{E_C\over
2\pi T})-2Re\,\psi(i{|\omega|\over 2\pi T})\right],\label{49a}\\
I(\omega)&=&D(\omega)\pi\omega\,\coth{({\omega\over 2T})},\label{49b}
\end{eqnarray}
$\psi$ is the digamma function and we have chosen a Lorentzian cutoff
function $D(\omega)=E_C^2/(\omega^2+E_C^2)$ with a band width given by the
charging energy $E_C$. For $E_C\gg|\omega-\mu_r|\gg T$, the self-energy
$\sigma(\omega)$ behaves marginally
\begin{equation}\label{51}
\sigma(\omega)=-\sum_r\alpha^r_0\left[2(\omega-\mu_r)\ln({E_C\over
|\omega-\mu_r|})+i\pi|\omega-\mu_r|\right].
\end{equation}
Since the imaginary part is linear in $\omega$, the broadening effects
manifests themselves in a very unusual way.

We will now anlayse two limiting cases, either $T\ll
|\omega-\mu_r|\ll E_C$ or $|\omega-\mu_r|\le T\ll E_C$. Furthermore,
we assume for simplicity and with regard to the applications we are
discussing in section 5.2 and 5.3 that all chemical potentials of the
reservoirs are zero.

In the first case, i.e. $T\ll|\omega|\ll E_C$, we can use Eq.(\ref{51})
and obtain from Eqs.(\ref{45})and (\ref{35}) for the spectral density
\begin{equation}\label{52}
A(\omega)\cong{|\omega|\over\Delta_0}\,\cdot\,{\tilde{\Delta}(\omega)\tilde{\alpha}
(\omega)\over
(\omega-\tilde{\Delta}(\omega))^2+(\pi\tilde{\Delta}(\omega)
\tilde{\alpha}(\omega))^2}\,\,,\,\,T\ll|\omega|\ll E_C
\end{equation}
where $\tilde{\Delta}$ and $\tilde{\alpha}$ are the renormalized
parameters
\begin{eqnarray}
\tilde{\Delta}(\omega)&=&{\Delta_0\over 1+2\alpha_0
\ln({E_C\over|\omega|})}\,\cdot\,
{1\over 1+\pi^2\tilde{\alpha}(\omega)^2}\,,\label{53}\\
\tilde{\alpha}(\omega)&=&{\alpha_0\over 1+2\alpha_0 \ln({E_C\over
|\omega|})}.\label{54}
\end{eqnarray}
Approximately, $A(\omega)$ will have a maximum value at
$\tilde{\Delta}_0=\tilde{\Delta}(\tilde{\Delta}_0)$ with a broadening
of the order of $\pi\tilde{\Delta}_0\tilde{\alpha}_0$ where
$\tilde{\alpha}_0=\tilde{\alpha}(\tilde{\Delta}_0)$.
For $\pi\tilde{\alpha}_0\ll1$ this broadening can be neglected and
we obtain
\begin{equation}\label{55}
\tilde{\Delta}_0={\Delta_0\over 1+2\alpha_0
\ln({E_C\over|\tilde{\Delta}_0|})}\qquad,\qquad\tilde{\alpha}_0=
{\alpha_0\over{1+2\alpha_0 \ln({E_C\over|\tilde{\Delta}_0|})}}
\end{equation}
for the renormalized gap and the renormalized dimensionless
conductance. This result coincides with the RG-analysis in
ref.\cite{Fal-Scho-Zim} and for small $\alpha_0$ where
$\tilde{\Delta}_0$ has been replaced by $\Delta_0$ on the r.h.s. of
Eq.(\ref{55}) with the results of ref.\cite{Mat}. This means
that the leading logarithmic terms are included in our
diagram series.

For $|\omega|\le T\ll E_C$, we can approximate $R(\omega)$ by
$2\omega\ln{({E_C\over 2\pi T})}$ and obtain
\begin{equation}\label{56}
A(\omega)\cong{\tilde{\Delta}_0\over
\Delta_0}\,\cdot\,{\tilde{\alpha}_0\,\omega\,
\coth{({\omega\over 2T})}\over
(\omega-\tilde{\Delta}_0)^2+(\pi\tilde{\alpha}_0\,\omega\,
\coth({\omega\over
2T}))^2}\quad,\quad|\omega|\le T\ll E_C
\end{equation}
where the renormalized parameters are now
\begin{equation}\label{57}
\tilde{\Delta}_0={\Delta_0\over 1+2\alpha_0 \ln({E_C\over 2\pi
T})}\quad,
\quad\tilde{\alpha}_0={\alpha_0\over 1+2\alpha_0 \ln ({E_C\over 2\pi
T})}
\end{equation}
which is the result (\ref{1}) quoted in the introduction. If
$\tilde{\Delta}_0\le T$, the spectral density (\ref{56}) has
approximately a maximum at $\tilde{\Delta}_0$ with a broadening of the
order of $\pi\tilde{\alpha}_0 T$. Again, for $\pi\tilde{\alpha}_0\ll1$
this broadening is small compared to $\tilde{\Delta}_0$. However, the
results (\ref{56}) and (\ref{57}) are independent of the value of
$\tilde{\Delta}_0$ and can be used always when we need the spectral
density only for $|\omega|\le T$. This is indeed the case for the
calculation of the differential conductance in linear response as will
be shown in section 5.3.

As long as we are only interested in the maximum point
$\tilde{\Delta}_0$ of the spectral density we can conclude that it is
given by (\ref{55}) for $\tilde{\Delta}_0\gg T$ and by (\ref{57}) for
$\tilde{\Delta}_0\le T$. This has also been proposed in
ref.\cite{Fal-Scho-Zim} (in addition, we can see here that the correct
replacement to get (\ref{57}) from (\ref{55}) is given by
$\tilde{\Delta}_0\rightarrow 2\pi T$ on the r.h.s. of
(\ref{55})). Furthermore, by using our solution (\ref{45}) for the spectral
density, we can also describe the crossover regime between
$\tilde{\Delta}_0\le T$ and $\tilde{\Delta}_0\gg T$ and we are able to
account for the finite broadening if $\pi\tilde{\alpha}_0$
approaches unity. The latter case is of interest,
since it is possible to achieve experimental parameters like
${E_C\over 2\pi T}\sim3$, $\alpha_0\sim0.5$ \cite{Est-pri} which gives
$\pi\tilde{\alpha}_0\sim0.75$ by using Eq.(\ref{57}).
On the other hand, when $\pi\as\ll 1$, the effects of energy
renormalization can become important when the temperature is very low.
E.g. for $\lnn=0.1$, $\pi\as\ll 1$ requires
$\alpha_0\sim 0.01$ according to (\ref{57}) which gives a temperature
of the order of $0.001 E_C$.  \\ \\
\noindent
{\bf\underline{5.2 Charge fluctuations in the single electron box}}\\
\\
In the equilibrium case where $\mu_r=0$, the SET-transistor becomes
equivalent to the single electron box. The average excess particle
number can be calculated from (\ref{9}) and (\ref{46b})
\begin{equation}\label{59}
\bar{N}=\int d\omega f(\omega) A(\omega).
\end{equation}
Within the classical approach given by Eq.(\ref{47}), one obtains
\begin{equation}\label{60}
\bar{N}^{cl}=f(\Delta_0)
\end{equation}
where the bare gap $\Delta_0=E_C(1-2C_gV_g)$ can also be expressed in
terms of the external gate voltage. Thus, $\bar{N}^{cl}(V_g)$ shows a
step at $V_g^{(0)}={C_g\over 2}$ (or $\Delta_0=0$) which is smeared by
temperature.

As can be seen from Fig.\ref{fig19} and Fig.\ref{fig20} there are
clear deviations from the classical result if $\alpha_0$ increases or
if the temperature T is decreasing. To estimate this, we have
neglected effects from the finite broadening, i.e.
$\pi\tilde{\alpha}_0\ll1$, and have assumed that
the energy of the ground state and the
first excited state lie symmetric (with distance ${\tilde{\Delta}_0
\over 2}$) around the degeneracy point
where $V(1)=V(0)={E_C\over 4}$ \cite{Fal-Scho-Zim}. In this case the
partition function reads
\begin{equation}\label{61}
Z\cong\,2\,e^{-{E_C\over 4T}}\,\cosh({\tilde{\Delta}_0\over 2T})
\end{equation}
and we obtain for the average excess particle number
$\bar{N}=n_x-T{\partial\over\partial\Delta_0}\ln Z$ (see Eq.(\ref{4}))
\begin{equation}\label{62}
\bar{N}\cong
{1\over2}(1-{\partial\tilde{\Delta}_0\over\partial\Delta_0}\,\tanh({\tilde
{\Delta}_0\over 2T})).
\end{equation}
 From (\ref{55}) and (\ref{57}) it follows for $\tilde{\Delta}_0\gg T$ as
well as for $\tilde{\Delta}_0\le T$ that ${\partial
\tilde{\Delta}_0\over
\partial\Delta_0}\cong{\tilde{\Delta}_0\over\Delta_0}$ where we have
used $\tilde{\alpha}_0\ll1$. Thus we find
\begin{eqnarray}
\bar{N}\cong{1\over2}(1-{1\over\nln}\tanh
({\Delta_0\over 2T(\nln)}))\quad&,\quad\ds\le T\quad,\quad \pi\as\ll 1&
\label{63}\\
\bar{N}\cong{1\over2}(1-{\Delta_0\over|\Delta_0|}{1\over 1+2\alpha_0
\ln({E_C\over |\Delta_0|})})\quad&,\quad\ds\gg T\quad,\quad \pi\as\ll 1&
\label{63aa}
\end{eqnarray}
For $\tilde{\Delta}_0\le T$ given by Eq.(\ref{57}), this means that
we get significant deviations from the
classical result (\ref{60}) if $2\alpha_0 \ln{({E_C\over 2\pi T})}\sim
1$. Thus by increasing $\alpha_0$ or decreasing T, quantum
fluctuations become more important. For $\ds\gg T$, Eq.(\ref{63aa})
coincides with the result obtained in \cite{Mat} for the $T=0$ case.

The slope at $\Delta_0=\ds=0$ is given by
\begin{equation}\label{63a}
{\partial \bar{N}\over\partial\Delta_0}|_{\Delta_0=0}=-{1\over
4T(1+2\alpha_0 \ln{({E_C\over 2\pi T})})^2}
\end{equation}
which differs from the classical result by the logarithmic term.

Thus, we conclude that the presence of coherent higher order
tunneling processes can be identified by
an anomalous temperature behaviour of the slope
of the Coulomb staircase at the degeneracy point.

Fig.\ref{fig21}a shows a comparison of the fits (\ref{63},\ref{63aa})
with the correct result obtained from
Eq.(\ref{59}) and (\ref{45}). Both line shapes agree quite well for
$\pi\tilde{\alpha}_0\ll 1$ where finite life-time broadening effects are not
important. Fig.\ref{fig21}b shows the same comparison for
$\pi\tilde{\alpha}_0\sim 1$ and we can see here clear differences for
$\ds\ge T$ whereas for $\ds\le T$ the approximation (\ref{63})
still seems to be reasonable.\\

\noindent
{\bf\underline{5.3 Conductance oscillations in the SET-transistor}}\\ \\
In the linear response regime we obtain for the current from (\ref{46c})
\begin{equation}\label{65}
I^{st}_R=-I^{st}_L=G\,(V_L-V_R)
\end{equation}
with the conductance G given by
\begin{equation}\label{66}
G=-{e^2\over h} 4\pi^2\int d\omega
{\alpha_R(\omega)\alpha_L(\omega)\over\alpha(\omega)}A(\omega)f^\prime(\omega).
\end{equation}
The derivative of the Fermi function restricts the integration variable
to the regime $\omega\le T$. Thus we can use (\ref{56})
and obtain
\begin{equation}\label{67}
G={e^2\over h}2\pi^2\,\,{\as^R\as^L\over\as}\int d\omega {\omega/T\over
\sinh(\omega/T)}\,\cdot\,{\as\,\omega\, \coth{({\omega\over
2T})}\over(\omega-\ds)^2+(\pi\as\,\omega\, \coth{({\omega\over 2T})})^2}
\end{equation}
where $\as=\as^R+\as^L$ and $\ds$ are given by Eq.(\ref{57}).

Two analytic results can be obtained from this formula. First, the
maximum conductance at $\Delta_0=\ds=0$ is given by
\begin{equation}\label{68}
G_{max}={e^2\over h} 2\pi{\as^R\as^L\over\as}\left[{\pi\over
2}-\arctan({(\pi\as)^2-1\over 2\as\pi})\right]
\end{equation}
and secondly, the integral of the conductance over the gap $\Delta_0$
(or equivalently over $e^2{C_g\over C}V_g$) is equal to
\begin{equation}\label{69}
\int d\Delta_0 \,\,G(\Delta_0)={e^2\over h}\pi^4\,\,{\alpha_0^R\alpha^L_0\over
\alpha_0}T.
\end{equation}
The broadening $\gamma$ of the conductance peak then follows from
\begin{equation}\label{70}
\gamma\cong{\int d\Delta_0 \,G(\Delta_0)\over G_{max}}={\pi^3\,T(\nln)\over
\pi-2\,\arctan({(\pi\as)^2-1\over2\as\pi})}.
\end{equation}
In the regime $\pi\as\ll 1$, (\ref{68}) and (\ref{70}) reduce to
\begin{eqnarray}
G_{max}&\cong&{e^2\over h}2\pi^2\,\,{\alpha_0^R\alpha^L_0\over
\alpha_0}\,\cdot\,{1\over \nln}\label{71a}\\
\gamma&\cong&{\pi^2\over 2}\,T\,(\nln)\label{71b}
\end{eqnarray}
These results lead to the following predictions.
In the regime $\lnn\ll1$, $G_{max}$
is constant and $\gamma$ is proportional to T. This is the classical
result for sequential tunneling. For $\lnn\sim 1$ and $\pi\as\ll 1$,
$G_{max}$ and $\gamma$ contain terms logarithmic in the temperature,
which are an indication for energy renormalization effects due to
higher-order tunneling processes. For $T\rightarrow 0$ we have
\begin{equation}\label{72}
G_{max}\sim{1\over \ln T}\qquad,\qquad \gamma\sim T\ln T
\end{equation}
i.e. the maximum value as well as the broadening go to zero. For
$\lnn\sim 1$ and $\pi\as\sim 1$, we have to account for energy
renormalization effects as well as effects from finite life-times
which lead to the complex formulas (\ref{68}) and
(\ref{70}). Fig.(\ref{fig22}) shows the conductance versus $\Delta_0$ for
several temperatures calculated numerically from formula
(\ref{66}). The result demonstrates the predicted behaviour.

For the conductance away from the degeneracy point
$\Delta_0=0$, we can make the following analytic analysis. For
$\ds\le T$ and $\pi\as\ll 1$, we can replace the last fraction in
Eq.(\ref{67}) by $\delta(\omega-\ds)$ since the broadening is of the
order $\pi\as T\ll T$ and the function ${\omega/T\over
\sinh(\omega/T)}$ varies on a scale of $\omega\sim T$. Thus we
obtain
\begin {equation}\label{73}
G={e^2\over h}2\pi^2\,\,{\as^R\as^L\over\as}\,\cdot\,{\ds/T\over \sinh(\ds/
T)}\qquad,\quad \ds\le T\quad,\quad\pi\as\ll 1
\end{equation}
which is the classical result but with renormalized parameters $\ds$
and $\as$.

For $\ds\gg T$ and $\pi\as\le 1$, we can replace the last
denominator in Eq.(\ref{67}) by $1/\ds^2$ since $\omega\le
T\ll\ds$ and $\pi\as\omega\le\pi\as T\le T\ll\ds$. This gives
\begin{equation}\label{74}
G={e^2\over h}{8\pi^4\over
3}\alpha_0^R\alpha^L_0\,({T\over\Delta_0})^2\qquad,\quad\ds\gg
T\quad,\quad\pi\as\le 1
\end{equation}
where the renormalization factor $(\nln)^{-1}$ has dropped out. Thus
in the Coulomb blockade regime we recover the usual expression of
inelastic electron cotunneling \cite{Ave-Odi,Ave-Naz} provided that
$\pi\as\le 1$ and $\ds\gg T$. Therefore, the influence of resonant
tunneling processes seems to be easier observable at resonance where
$\ds\le T$. Here we expect significant deviations from sequential
tunneling already in the regime $\lnn\sim 1$ and effects from finite
life-times (see Eq.(\ref{68})) for $\pi\as\sim 1$.

Finally, we will compare our result (\ref{67}) with the one obtained in
\cite{Ave} which also coincides with other approaches
\cite{Kor-etal,Naz,Laf-Est,Pas-etal}. It is given by
\begin{equation}\label{75}
G={e^2\over h}2\pi^2\,{\alpha_0^R\alpha_0^L\over \alpha_0}\int d\omega
{\omega/T\over \sinh(\omega/T)}\,\cdot\,{\alpha_0\,\omega\, \coth({\omega\over
2T})\over (\omega-\Delta_0)^2+(\pi \alpha_0\Delta_0
\coth({\Delta_0\over 2T}))^2}.
\end{equation}
Here the energy renormalization effects in $\alpha_0$ and $\Delta_0$
have been neglected and the integration variable $\omega$ is replaced
by $\Delta_0$ in the broadening part of the last denominator. This
corresponds to the introduction of a constant and finite life-time
into the usual expressions of inelastic electron cotunnneling
\cite{Ave-Odi,Ave-Naz} which regularizes the integrals. This is only
justified in the regime $\lnn\ll 1$ where the renormalization factor
$(\nln)^{-1}$ and the broadening $\pi\as\omega \coth({\omega\over
2T})\sim\pi\alpha_0 T\ll T$ are unimportant. Thus the replacement
$\omega\rightarrow\Delta_0$ within the broadening doesn't change the
result for sequential tunneling significantly ($\Delta_0\le T$)
and for $\Delta_0\gg T$ Eq.(\ref{75}) leads to the result for
electron cotunneling since the numerator of the last fraction of
Eq.(\ref{67}) has not been changed. However, for $\lnn\sim 1$ or
$\pi\as\sim 1$, Eq.(\ref{75}) can no longer be used at resonance as is
demonstrated in Fig.(\ref{fig23}).\\ \\ \\ \\

\section{Conclusions}
In this paper, we have aimed at presenting a detailed theory of
quantum fluctuation effects in transport through small metallic
islands with strong Coulomb interaction. With the help of a
diagrammatic technique in real-time space, we have identified and
evaluated the contribution of correlated higher-order tunneling
events.
Assuming a wide junction with many transverse channels, we have allowed
different electrons to tunnel an arbitrary number of times coherently
between the leads and the metallic island.
Using the two charge state approximation we have included
in a closed analytic form sequential
tunneling, inelastic electron cotunneling and resonant tunneling
processes.

 From a theoretical point of view it has turned out that the effects of
quantum fluctuations
can be understood very clearly by investigating the spectral density
which describes the charge excitations of the system. In the classical
regime, the spectral density has a sharp maximum at the gap energy
$\Delta_0$ which is the difference of the Coulomb energies of two adjacent
charge states. In the quantum regime, the maximum point renormalizes
to $\ds$ and we obtain a finite broadening which can be estimated to
be of the order of $\pi\as \,max\{\ds,T\}$ where $\as$ is the renormalized
dimensionless conductance of a single barrier. Both features together
with the complete form of the spectral density is described by the
real and imaginary part of the self-energy $\sigma(\omega)$. It
contains an anomalous dependence on energy and temperature which leads
to a variety of unexpected features in the line shapes of several
experimental quantities.

To estimate the experimental consequences of quantum fluctuations we
have calculated the average charge of the single electron box and the
linear conductance of the SET-transistor as function of the gap energy
$\Delta_0$ or equivalently the external gate voltage. For the average
charge we have compared our results to previous investigations
\cite{Mat,Fal-Scho-Zim} where renormalization group techniques have
been used to study the zero temperature case. Using the temperature as
a cutoff one can calculate the average charge from these theories
in the two limiting case $\ds\gg T$ or $\ds\le T$. These results agree
with our solution in the case $\pi\as\ll 1$, i.e. in the regime where
finite life-time effects are not important. For $\pi\as\sim 1$ there
are significant deviations at least for $\ds\ge T$. Furthermore our
complete solution is capable of describing the complete crossover from
$\ds\le T$ to $\ds\gg T$ and shows that the temperature has to be
introduced into the renormalization of the system parameters by the
replacement $\ds\rightarrow 2\pi T$.

For the conductance in the linear response regime we have seen that
the classical description of sequential tunneling near the resonance
($\ds\le T$) is only valid for $\lnn\ll 1$ and
$\pi\as=\pi\alpha_0/(\nln)\ll 1$. Nowadays experiments are not
restricted to this regime. Therefore we expect that resonant
tunneling of coherent
higher order tunneling processes should be observable by a measurement
of the line shape of the conductance peaks as function of $\alpha_0$
and temperature.
The renormalization of the system parameters $\Delta_0$ and $\alpha_0$
is important for $\lnn\sim 1$ and leads to an anomalous logarithmic
temperature dependence of the conductance peak and the broadening.
For $T\rightarrow 0$ the conductance maximum decrease like
$1/\ln T$ and the broadening increases proportional to $T\ln T$.
Furthermore, for $\pi\as\sim 1$, the influence of finite life-times
becomes very important and leads to a very significant flattening of the
Coulomb oscillations. Both effects are important and we have seen that
it is very difficult to separate them at realistic temperatures.

Finally our approach describes also the conductance in the
Coulomb-blockade regime where transport is dominated by inelastic
electron cotunneling. Our analytic formulas give the correct crossover
from resonant tunneling at the degeneracy point to inelastic
cotunneling. We have seen that cotunneling persists in the regime
$\ds\gg T$ and $\pi\as\le 1$. For $\pi\as\gg 1$ cotunneling as well as
the validity of our approach will break down. In this regime the
charge will no longer be well defined and one should go beyond the two
charge state approximation.

\acknowledgments
We would like to acknowledge many stimulating and helpful discussions
with S.E. Barnes, D. Esteve, H. Grabert, P. Falci, K.A. Matveev,
Y.V. Nazarov, A. Schmid, P. W\"olfle, A.D. Zaikin and G. Zimanyi.
This work was supported by the Swiss National Science
Foundation (H.S.) and by the 'Deutsche Forschungsgemeinschaft' as part of
'Sonderforschungsbereich 195'.


\begin{figure}
\caption{
Equivalent circuit for the SET-transistor.
}
\label{fig1}
\end{figure}

\begin{figure}
\caption{
Example of a diagram in time space for the probability $P_N(t)$. Time
is increasing from left to right. The dots represent the vertices
which are connected by horizontal lines (propagators), solid lines
(leads) or wiggly lines (island). $\epsilon_j, E_j$ are the
corresponding energies, $r_j$
are the lead indices and N denotes the excess particle number.
}
\label{fig2}
\end{figure}

\begin{figure}
\caption{
Graphical representation of (a) the current $I_r$ through lead r and
(b) the correlation functions $C^>, C^<$ in time space. The dots
indicate the external vertices. Further internal vertices together with
their connections are not indicated.
}
\label{fig3}
\end{figure}

\begin{figure}
\caption{
Graphical determination of the energy denominators. The energies of
all lines contribute to $\Delta E$ which are cut by an auxiliary
vertical line.
}
\label{fig4}
\end{figure}

\begin{figure}
\caption{
If an auxiliary vertical line cuts the virtual line connecting the
external vertices, the energy $\omega$ contributes to the energy
denominators.
}
\label{fig5}
\end{figure}

\begin{figure}
\caption{
Graphical representation of the correlation functions in energy space.
}
\label{fig6}
\end{figure}

\begin{figure}
\caption{
Self-consistent equation for the stationary probability
$P_N^{st}$. The self-energy $\Sigma$ denotes the sum of all
irreducible diagrams which can not be cut into two parts by an arbitrary
vertical line.
}
\label{fig7}
\end{figure}

\begin{figure}
\caption{
Examples for irreducible diagrams. An arbitrary vertical line will
always cut through some reservoir line.
}
\label{fig8}
\end{figure}

\begin{figure}
\caption{
First order diagrams for $\Sigma$ which correspond to the classical
transition rates.
}
\label{fig9}
\end{figure}

\begin{figure}
\caption{
Replacement of a loop consisting of one lead and one island line by a
single solid line. The arrow direction is the same as of the lead line.
}
\label{fig10}
\end{figure}

\begin{figure}
\caption{
Second order diagram for $\Sigma_{00}$ which contributes to
inelastic electron cotunneling when the two leads $r$ and $r^\prime$
are different.
}
\label{fig11}
\end{figure}

\begin{figure}
\caption{
Higher order diagrams which contribute to resonant tunneling.
}
\label{fig12}
\end{figure}

\begin{figure}
\caption{
Example of a diagram which is taken into account. All vertical lines
cut at most through two solid lines. The cut through the two
horizontal lines is trivial and is not counted.
}
\label{fig13}
\end{figure}

\begin{figure}
\caption{
Graphical representation of the self-energy $\Sigma_{N1}$ within our
approximation.
}
\label{fig14}
\end{figure}

\begin{figure}
\caption{
Self-consistent equation for $\phi^r_N(\omega)$.
}
\label{fig15}
\end{figure}

\begin{figure}
\caption{
Definition of $\Pi$. To each energy denominator one has to add
$\omega$ in order to obtain $\Pi(\omega)$.
}
\label{fig16}
\end{figure}

\begin{figure}
\caption{
Graphical representation of the relation between the current and the
correlation functions. Here the line connecting the external vertices
is a real one. All the other internal vertices are not indicated.
}
\label{fig17}
\end{figure}

\begin{figure}
\caption{
Graphical representation of (a) $C^>(\omega)$ and (b) $C^<(\omega)$
within our approximation.
}
\label{fig18}
\end{figure}

\begin{figure}
\caption{
Average charge as function of the gap energy for different
values of $\alpha_0$. $E_C=1\,,\, T=0.01$ and (a) $\alpha_0=0\,$, (b)
$\alpha_0=0.01$ and (c) $\alpha_0=0.1$. Curve (a) is the Fermi
distribution function which corresponds to the classical result.
For increasing $\alpha_0$ the deviations become more significant.
}
\label{fig19}
\end{figure}

\begin{figure}
\caption{
Average charge as function of the gap energy at zero temperature. At
finite $\alpha_0=0.03$ there are clear deviations from a pure step
function.
}
\label{fig20}
\end{figure}

\begin{figure}
\caption{
Average charge as function of the gap energy using (1) the correct
result Eq.(67), (2) the fit (71) and (3) the fit (72).
$E_C=1\,,\, T=0.01$ and (a)
$\alpha_0=0.02$ and (b) $\alpha_0=0.2$. When finite life-time
broadening effects set on in (b) the deviations grow.
}
\label{fig21}
\end{figure}

\begin{figure}
\caption{
Conductance as function of the gap energy for various
temperatures. $E_C=1\,,\,\alpha_0=0.02$ and (1) $T=0.05\,$, (2) $T=0.02\,$, (3)
$T=0.005$ and (4) $T=0.0005$.
}
\label{fig22}
\end{figure}

\begin{figure}
\caption{
Conductance as function of the gap energy using (1) the correct result
(76), (2) the fit (86) and (3) the fit (84). $E_C=1$ and (a)
$T=0.001\,,\,\alpha_0=0.02$ and (b) $T=0.05\,,\,\alpha_0=0.6$.
In both cases are clear differences to the classical result (86). (a)
can be described by renormalized parameters given by (84) whereas in
(b) finite life-time broadening effects dominate.
}
\label{fig23}
\end{figure}

\end{document}